\documentclass[12pt,preprint]{aastex}

\newcounter{address}
\newcommand{\latin}[1]{{#1}}
\newcommand{\ie}{\latin{i.e.}}
\newcommand{\eg}{\latin{e.g.}}

\newcommand{\Halpha}{\ensuremath{\mathrm{H}\alpha}}
\newcommand{\Dcl} {\ensuremath{D_\mathrm{cl}}}
\newcommand{\Rvir} {\ensuremath{R_\mathrm{vir}}}
\newcommand{\N}{\ensuremath{N_\mathrm{gal}}}

\begin{document}
\title{ The asymmetric relations among galaxy color, structure, and environment }
\author{
  Alejandro~D.~Quintero\altaffilmark{\ref{Steward}},
  Andreas~A.~Berlind\altaffilmark{\ref{NYU}},
  Michael~R.~Blanton\altaffilmark{\ref{NYU}},
  David~W.~Hogg\altaffilmark{\ref{NYU}} }

\setcounter{address}{1}
\altaffiltext{\theaddress}{\stepcounter{address}\label{Steward}
Steward Observatory, 933 N. Cherry Ave., Tucson, AZ 85721}
\altaffiltext{\theaddress}{\stepcounter{address}\label{NYU}
Center for Cosmology and Particle Physics, Department of Physics, New
York University, 4 Washington Place, New York, NY 10003}


\begin{abstract}
We investigate the dependences of galaxy star-formation history and
galaxy morphology on environment, using color and \Halpha\ equivalent
width as star-formation history indicators, using concentration and
central surface brightness as morphological indicators, and using
clustocentric distance as an environment indicator.  Clustocentric
distance has the virtue that it can be measured with very high
precision over a large dynamic range.  We find the following asymmetry
between morphological and star-formation history parameters:
star-formation history parameters relate directly to the clustocentric
distance while morphological parameters relate to the clustocentric
distance only \emph{indirectly} through their relationships with
star-formation history.This asymmetry has important implications for
the role that environment plays in shaping galaxy properties and it
places strong constraints on theoretical models of galaxy formation.
Current semi-analytic models do not reproduce this effect.
\end{abstract}

\keywords{
    galaxies: clustering
    ---
    galaxies: clusters: general
    ---
    galaxies: evolution
    ---
    galaxies: fundamental parameters
    ---
    galaxies: statistics
    ---
    galaxies: stellar content
}

\section{Introduction}

It has been known for quite some time that the statistical properties
of galaxies are closely related to their surrounding environments.
Much of previous environment-related work focused on the relationship
between morphology and environment (\eg, \citealt{hubble36a,
oemler74a, dressler80a, postman84a}; or the more recent work of
\citealt{hermit96a, guzzo97a, giuricin01a, trujillo02a}).  All these
works find that bulge-dominated galaxies are more strongly clustered
than disk-dominated galaxies.  Since spectroscopic and photometric
properties of galaxies are strongly correlated with morphology, it is
not surprising to find that the spectroscopic and photometric
properties of galaxies are also functions of environment (\eg,
\citealt{kennicutt83a, hashimoto98a, balogh01a, martinez02a, lewis02a,
norberg02a, hogg03b, hogg04a, kauffmann04a, zehavi05b}).

Since regions of the Universe with different densities evolve at
different rates, we expect these environment dependencies to contain
crucial information about galaxy formation and evolution.  Considering
that all of the observed statistical properties of galaxies appear to
be interdependent, it is important to ask which properties are
correlated with environment independently of the others; \ie, which
environment relationships are fundamental and which ones are merely
the products of other relations.

\cite{blanton05b} found that, out of color, luminosity, surface
brightness, and radial concentration, the color and luminosity of a
galaxy appear to be the only properties that are directly related to
the local overdensity.  Surface brightness and concentration appear to
be related to the environment only through their relationships with
color and luminosity.  The main limitations in this study were caused
by their environment indicator, local overdensity, since it has
relatively low signal-to-noise and it cannot resolve scales smaller
than $\sim1$\,Mpc.  In a similar study, \cite{christlein05a} used the
clustocentric distance as an environment indicator found that there is
a residual dependence of ongoing star formation on environment among
galaxies of similar morphology, stellar mass, and mean stellar age.
This study was limited by a relatively small sample size (1,637
galaxies in 6 clusters) as well as by the star formation indicator,
OII, which is sensitive to dust and metallicity.

In this short paper, we build upon previous studies in several ways.
We conduct a study similar to \cite{blanton05b}, except that we use
the clustocentric distance as our environment indicator.  The
clustocentric distance---the distance to the nearest galaxy cluster
center---is a fundamental and precisely measurable environment
indicator; indeed it can be measured at much higher signal-to-noise
than estimates of overdensity on fixed scales.  Thus, we probe the
findings of \cite{blanton05b} on much smaller scales and at much
higher signal-to-noise.  We use the \cite{berlind06a} cluster catalog
that was created with an algorithm that made no use of galaxy colors,
so our galaxy colors are unbiased even in the cluster cores.  The
cluster catalog was designed to recover groups of galaxies that are
bound in the same underlying dark matter halo.  In order to determine
whether our results can be easily understood within our current
understanding of galaxy formation and evolution, we compare our
observational results from the Sloan Digital Sky Survey to analogous
results from the semi-analytic Millennium simulation of galaxy
formation \citep{croton06a}.

In what follows, a cosmological world model with
$(\Omega_\mathrm{M},\Omega_\mathrm{\Lambda})=(0.3,0.7)$ is adopted,
the Hubble constant is parameterized as
$H_0\equiv100\,h~\mathrm{km\,s^{-1}\,Mpc^{-1}}$, and for the purposes of
calculating distances and volumes $h=1$ except where otherwise
noted \citep[\eg,][]{hogg99cosm}.

\section{Data and Analysis}

\subsection{Observations}

The SDSS has taken $ugriz$ CCD imaging and spectroscopy of many $10^5$
galaxies at $r<17.77~\mathrm{mag}$
\citep[\eg,][]{gunn98a,york00a,stoughton02a}.  All the data
processing, including astrometry \citep{pier03a}, source
identification, deblending and photometry \citep{lupton01a},
calibration \citep{fukugita96a,smith02a}, spectroscopic target
selection \citep{eisenstein01a,strauss02a,richards02a}, spectroscopic
fiber placement \citep{blanton03a}, spectral data reduction and
analysis (Schlegel \& Burles, in preparation, Schlegel in preparation)
are performed with automated SDSS software.

Galaxy colors are computed in fixed bandpasses, using Galactic
extinction corrections \citep{schlegel98a} and $K$ corrections
\citep[computed with \texttt{kcorrect v3\_2};][]{blanton03b}.  They
are $K$ corrected to the redshift $z=0$ observed bandpasses so they
can be directly compared to the rest-frame SDSS $ugriz$ outputs of the
Millennium simulation \citep{croton06a}.

For the purposes of computing large-scale structure statistics, we
have assembled a complete subsample of SDSS galaxies known as the NYU
LSS \texttt{sample14}.  This subsample is described elsewhere
\citep{blanton05a}; it is selected to have a well-defined window
function and magnitude limit.  In addition, the galaxies in the sample
used here were selected to have apparent magnitudes in the range
$14.5<r<17.77~\mathrm{mag}$, redshift in the range $0.015<z<0.068$,
and absolute magnitude $M_{^{0.1}i}>-24~\mathrm{mag}$.  These cuts
left 52,485 galaxies.

A seeing-convolved S\'ersic model is fit to the azimuthally averaged 
radial profile of every galaxy in the observed-frame $i$ band, as
described elsewhere \citep{blanton03d,strateva01a}.  The S\'ersic
model has surface brightness $I$ related to angular radius $r$ by
$I\propto \exp[-(r/r_0)^{(1/n)}]$, so the parameter $n$ (S\'ersic
index) is a measure of radial concentration (seeing-corrected).  At
$n=1$ the profile is exponential, and at $n=4$ the profile is
de~Vaucouleurs.  In the fits shown here, values in the range
$0.5<n<5.5$ were allowed.

To every best-fit S\'ersic profile, the \citet{petrosian76a}
photometry technique is applied, with the same parameters as used in
the SDSS survey.  This supplies seeing-corrected Petrosian magnitudes
and radii.  A $K$-corrected surface-brightness $\mu_{^{0.0}i}$ in the
$^{0.0}i$ band is computed by dividing half the $K$-corrected
Petrosian light by the area of the Petrosian half-light circle.

The \Halpha\ line flux is measured in a 20~\AA\ width interval
centered on the line.  Before the flux is computed, a best-fit model
consisting of scaled old-galaxy and A-star spectra \citep{quintero04a}
is scaled to have the same flux continuum as the data in the vicinity
of the emission line and subtracted to leave a continuum-subtracted
line spectrum.  This method fairly accurately models the \Halpha\
absorption trough in the continuum, although in detail it leaves small
negative residuals.  The flux is converted to a rest-frame EW with a
continuum found by taking the inverse-variance-weighted average of two
sections of the spectrum about 150~\AA\ in size and on either side of
the emission line.  Further details are presented elsewhere
\citep{quintero04a}.

A caveat to this analysis is that the 3~arcsec diameter spectroscopic
fibers of the SDSS spectrographs do not obtain all of each galaxy's
light because at redshifts of $0.015<z<0.068$ they represent apertures
of between $0.6$ and $2.7~h^{-1}\,\mathrm{kpc}$ diameter.
Consequently the integrity of our \Halpha\ EW measurement is a
function of galaxy size, inclination, and morphology.  We have looked
at variations in our \Halpha\ EW results as a function of redshift and
found that the quantitative results differ but our qualitative results
remain the same.

We use the group and cluster catalog described in \cite{berlind06a}.
The catalog is obtained from a volume-limited sample of galaxies that
is complete down to an $^{0.1}r$ band absolute magnitude of
$M_r<-19$~mag and goes out to a redshift of 0.068.  Groups are
identified using a friends-of-friends algorithm (see e.g., Geller \&
Huchra 1983; Davis et al 1985) with perpendicular and line-of-sight
linking lengths equal to 0.14 and 0.75 times the mean inter-galaxy
separation, respectively.  These parameters were chosen with the help
of mock galaxy catalogs to produce galaxy groups that most closely
resemble galaxy systems that occupy the same dark matter halos.  The
resulting catalog contains 985 systems of richness $\N \geq 5$
galaxies with an r-band absolute magnitude limit of $M_r<-19$~mag.
For consistency, we call these objects ``clusters''.  Note that,
unlike some other catalogs, galaxy colors are \emph{not} used in
cluster identification.

Berlind et al. (2006b, in preparation) calculate rough mass estimates
for the clusters using the cluster luminosity function (where
luminosity is defined as the total luminosity in $M_r<-19$~mag
galaxies in the cluster) and assuming a monotonic relation between a
cluster's luminosity and the mass of its underlying dark matter halo.
By matching the measured space density of clusters to the theoretical
space density of dark matter halos (given the concordance cosmological
model and a standard halo mass function), they assign a virial halo
mass to each cluster luminosity.  The masses derived in this way
ignore the scatter in mass at fixed cluster luminosity and are only
meant to be approximate.  The median mass estimate for the clusters
used here is $3\times 10^{13}$ solar masses.  Each cluster has an
associated ``virial radius'' of
\begin{equation}
\Rvir =
\left(\frac{3}{4\,\pi}\frac{M}{200\,\rho_{o}}\right)^{\frac{1}{3}}
\quad ,
\end{equation}
where $M$ is the estimated mass of the cluster and $\rho_{o}$ is the
current mean density of the Universe.  By this method, the median
estimated virial radius for the clusters used here is $\sim1$~Mpc.
Note that this method for determination of the virial radii is very
different from that employed by other investigators.  Some have used a
quasi-empirical formula based on velocity dispersion \citep{gomez03a,
christlein05a}, others have assumed that cluster mass is directly
proportional to richness \citep{lewis02a}; in general these methods
differ substantially, and produce cluster catalogs with very different
mass functions.

We use the cluster centers given by \citet{berlind06a}, which are
computed as the mean of the member galaxy positions.  We then
calculate the transverse projected clustocentric distance \Dcl\ from
each galaxy to its nearest cluster center on the sky within
$\pm1000~\mathrm{km\,s^{-1}}$ in radial velocity.

\subsection{Simulations}

To better understand our observations, we do a similar analysis for
the Millennium simulation galaxy catalog.  The Millennium simulation
is a large cosmological N-body simulation of $2160^{3}$ dark matter
particles in a $500^{3}~\,h^{-3}\mathrm{Mpc^{3}}$ box.  The formation
of galaxies is simulated by using a semi-analytical model whose
parameters and assumptions are tweaked in order to reproduce the joint
luminosity-color, morphology, gas mass, and central black hole mass
distributions of low-redshift galaxies.  For a more detailed
description, see \cite{croton06a}.

The publicly available output of this simulation \citep{croton06a}
is a catalog of $\sim9$ million galaxies that contains positions,
velocities, bulge and total SDSS-band magnitudes, bulge and total
masses, gas masses (cold, hot, and ejected), black hole masses, and
star formation rates.

From this catalog we use the ratio of bulge-to-total luminosity as a
morphology indicator.  This $B/T$ value is an analog to our observed
concentration and surface brightness.  Similarly, we use the star
formation rate SFR and SDSS-band $^{0.0}[g-r]$ color from this catalog
as star-formation history indicators; these are analogs to our
\Halpha\ EW and $^{0.0}[g-r]$ color measurements, respectively.

Before running the cluster-finding algorithm \citep{berlind06a}, we
distort the Millennium coordinates to mimic redshift-space distortions
by changing the output coordinates of each galaxy from $[(x), (y),
(z)]$ to $[(x), (y), (z+\frac{v_{z}}{H_0})]$.  We then run the same
friends-of-friends algorithm that we did on the data, using the
perpendicular linking length in the $x-$ and $y-$directions and the
line-of-sight linking length in the $z$-direction.  After creating
this cluster catalog we calculate each galaxy's clustocentric distance
using a similar algorithm we use on the data: We calculate the
transverse projected clustocentric distance \Dcl\ from each galaxy to
its nearest cluster center on the $x-y$ plane within
$\pm10\,\mathrm{Mpc}$ in the $z$-direction.

\section{Results}

\begin{figure*}
\begin{center}
\resizebox{!}{5.5in}{\includegraphics{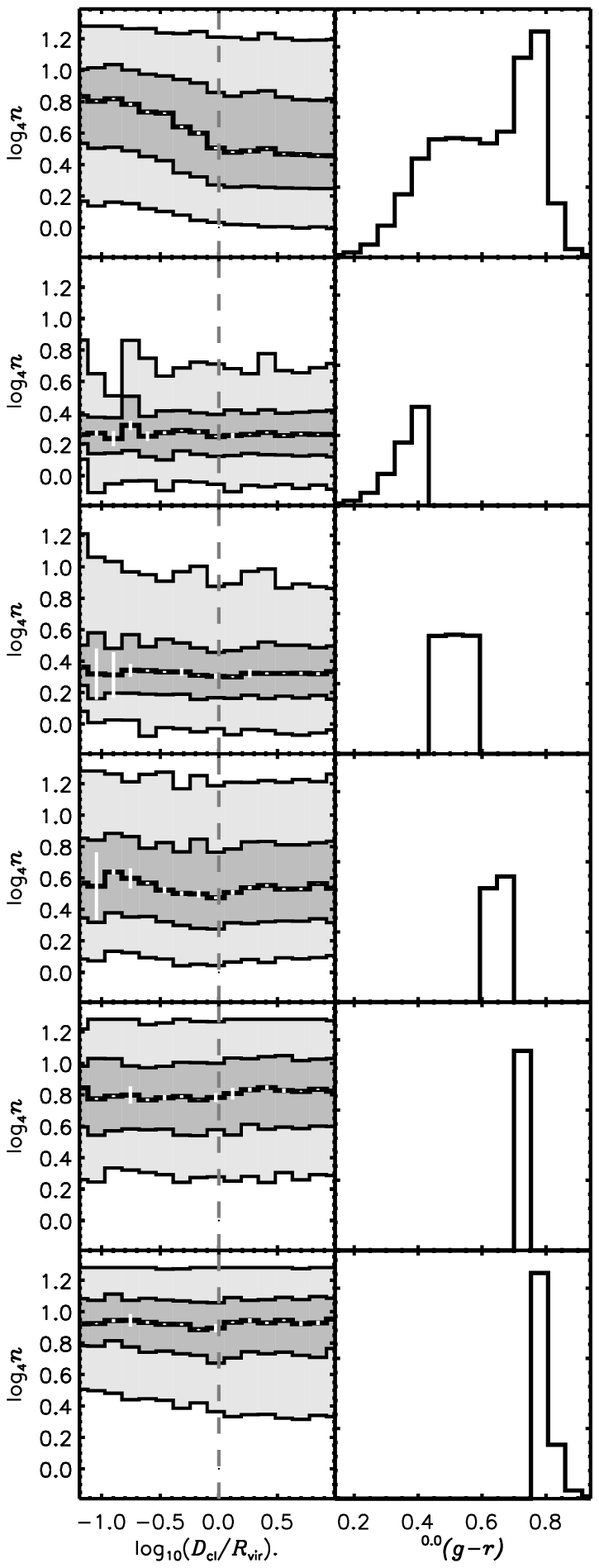}}%
\resizebox{!}{5.5in}{\includegraphics{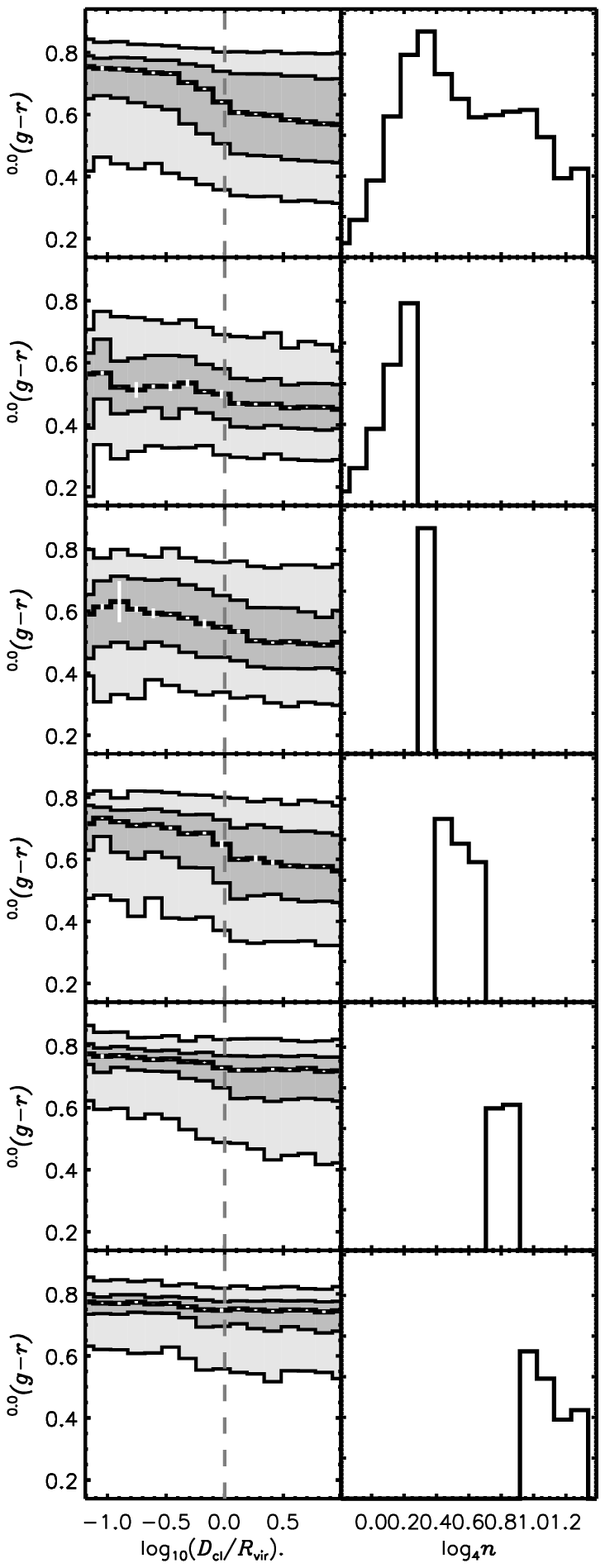}}
\end{center}
\caption{The dependence of radial concentration $n$ (left grid) and
$^{0.0}[g-r]$ color (right grid) on clustocentric distance \Dcl.  For
the $2\times6$~grid on the left, the top-left panel shows how the 5,
25, 50 (in bold), 75, and 95 percentiles of the distribution of
concentration $n$ (S\'ersic index) depend on clustocentric distance
\Dcl\ for the entire sample.  The top-right panel of this grid shows
the $^{0.0}[g-r]$ color distribution of the entire sample.  The errors
in the 50 percentile, calculated using the jackknife method (10
jackknife trials in each of which a contiguous 1/10 of the survey
footprint is dropped) are overplotted in white (they are tiny).  The
subsequent rows show the same but for narrow color subsamples (color
quintiles).  The $2\times6$~grid on the right is very similar but with
the $n$ and $^{0.0}[g-r]$ properties interchanged.  The left
$2\times6$~grid shows that in color subsamples, there is no residual
dependence of concentration on environment, while the right
$2\times6$~grid shows that in concentration subsamples, there
\emph{is} a residual dependence of color on environment.  This
asymmetry is the principal result of this paper.
\label{fig:n_dont_matter}}
\end{figure*}

\begin{figure*}
\begin{center}
\resizebox{!}{5.5in}{\includegraphics{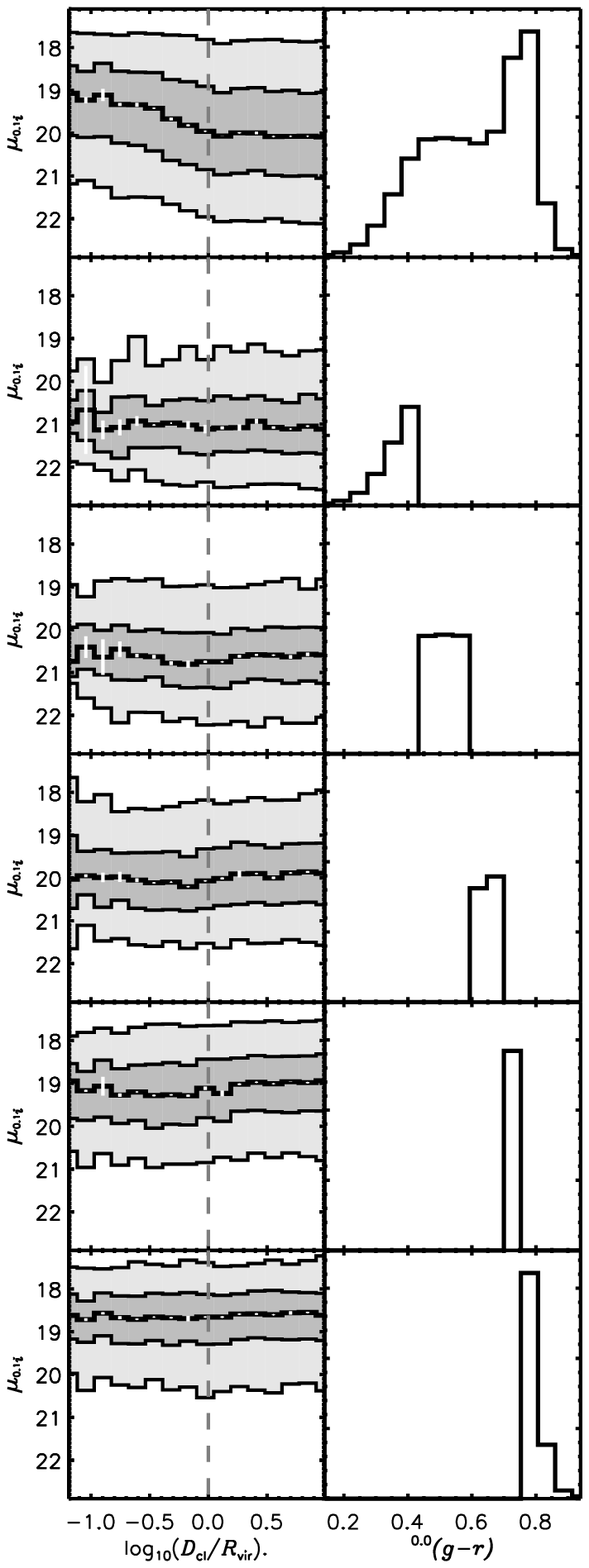}}%
\resizebox{!}{5.5in}{\includegraphics{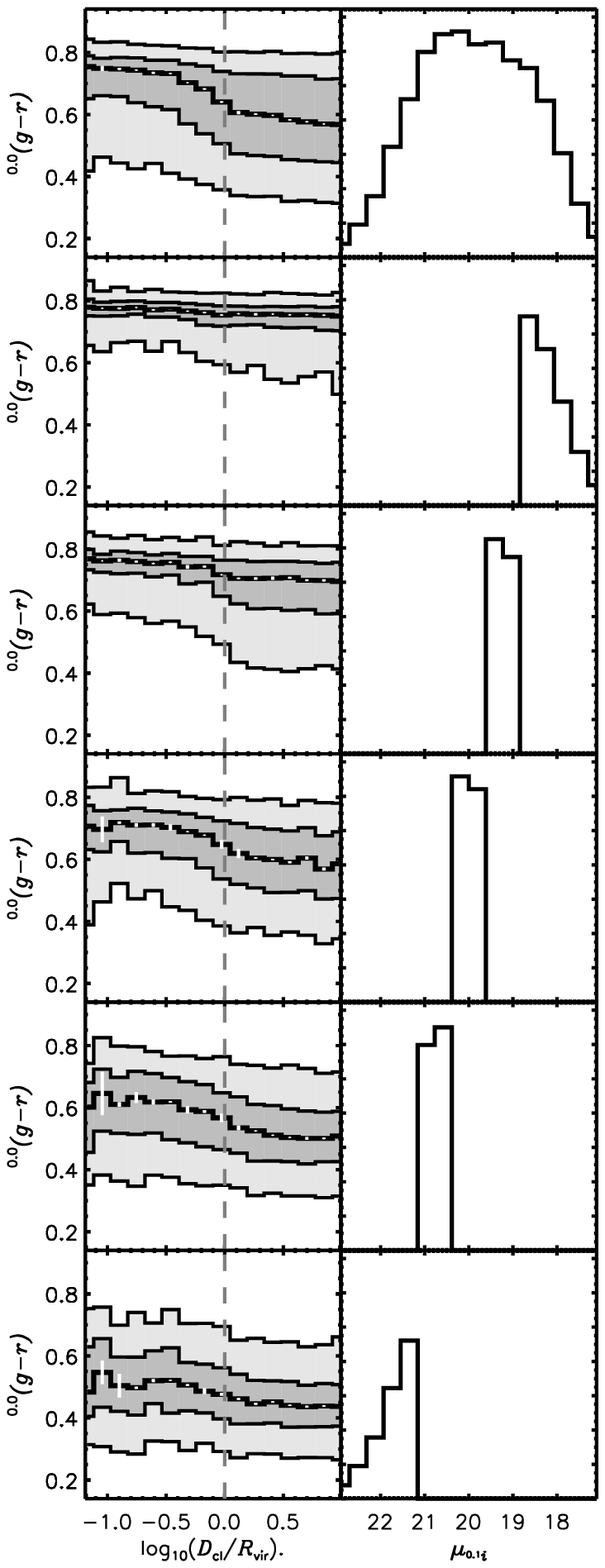}}
\end{center}
\caption{Similar to Figure~\ref{fig:n_dont_matter}, except replacing
the radial concentration $n$ with surface brightness $\mu_{^{0.0}i}$.
Notice the asymmetry seen in Figure~\ref{fig:n_dont_matter} also
occurs here.
\label{fig:mu_dont_matter}}
\end{figure*}

Figure~\ref{fig:n_dont_matter} shows how radial concentration and
color relate to clustocentric distance within narrow subsamples of
each other.  For the $2\times6$~grid on the left, the top-left panel
shows how the 5, 25, 50 (in bold), 75, and 95 percentiles of the
concentration distribution vary with clustocentric distance for the
entire sample.  The color distribution for the entire sample is shown
in the right panel.  The gradient in the concentration percentiles
show that galaxies with small clustocentric distances tend to be more
concentrated.  The subsequent rows show the same relation but for
narrow color subsamples whose color distribution is shown in the right
panels.  Notice the gradient seen in the top panel nearly vanishes for
the narrow color subsamples below.  The absence of a concentration
gradient for these subsamples shows that at fixed color, there is no
dependence of concentration on environment.  The right $2\times6$~grid
shows the same plot but with concentration and color interchanged.
Notice that a color gradient remains when looking at narrow
concentration subsamples.  This shows that at fixed concentration,
there \emph{is} a residual dependence of color on environment.  The
asymmetry seen in this figure is also seen in
Figure~\ref{fig:mu_dont_matter} which uses surface brightness, instead
of concentration, as a morphology tracer.  This asymmetry is the
principal result of this paper.  This asymmetry is similar to that
found by \cite{blanton05b}, but it appears here at much higher
significance and over a larger dynamic range.

\begin{figure*}
\begin{center}
\resizebox{!}{5.5in}{\includegraphics{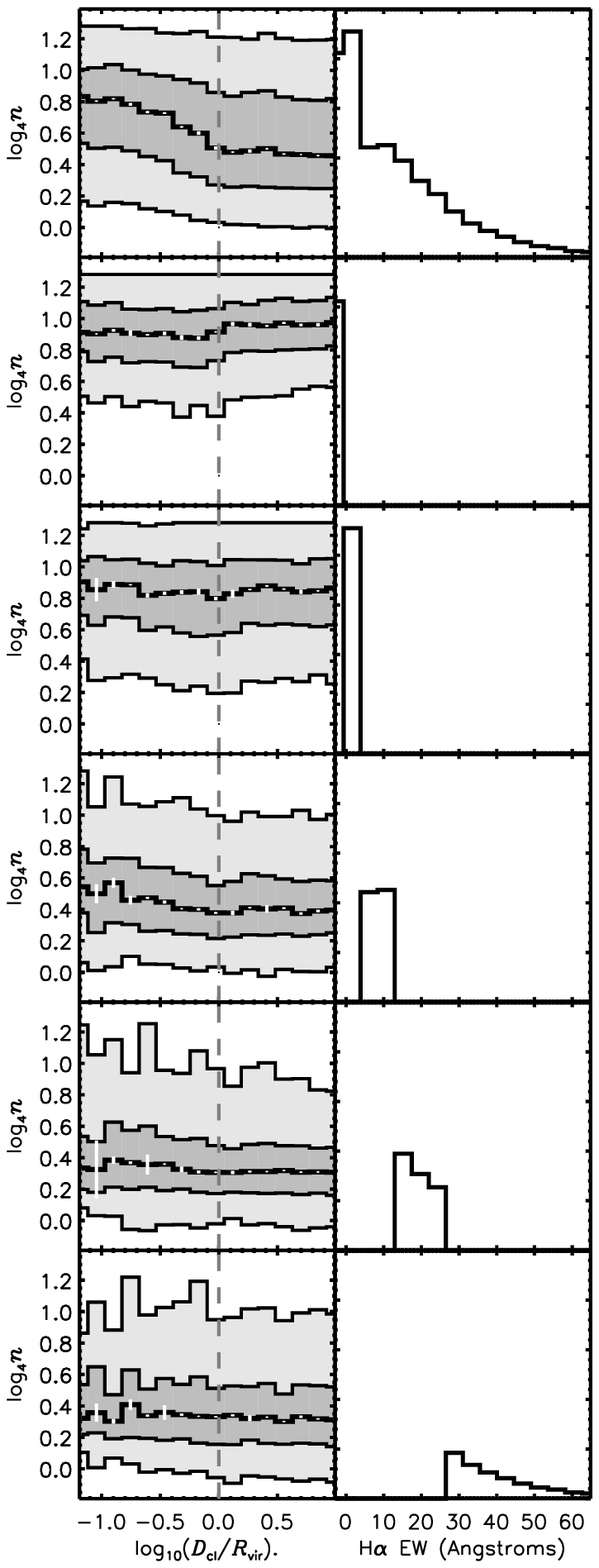}}%
\resizebox{!}{5.5in}{\includegraphics{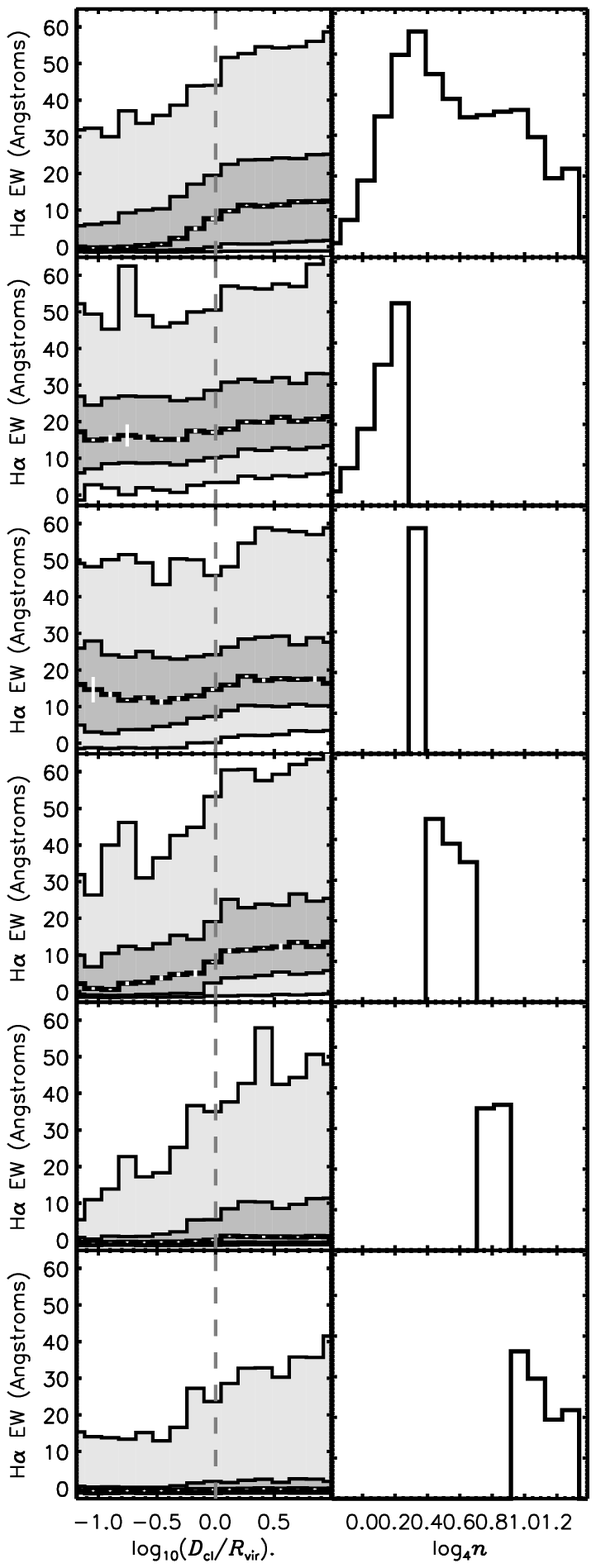}}
\end{center}
\caption{Similar to Figure~\ref{fig:n_dont_matter}, except replacing
the $^{0.0}[g-r]$ color with our other star-formation history
indicator: \Halpha\ emission line equivalent width (\Halpha\ EW).
Notice the asymmetry seen in Figure~\ref{fig:n_dont_matter} also
occurs here.
\label{fig:ha_matter}}
\end{figure*}

Figure~\ref{fig:ha_matter} also shows the same asymmetry shown in
Figure~\ref{fig:n_dont_matter} but using \Halpha\ EW, instead of
color, as our star-formation-history parameter.  Note that the right
$2\times6$~grid agrees with the findings of \cite{christlein05a}.

\begin{figure*}
\begin{center}
\resizebox{2in}{!}{\includegraphics{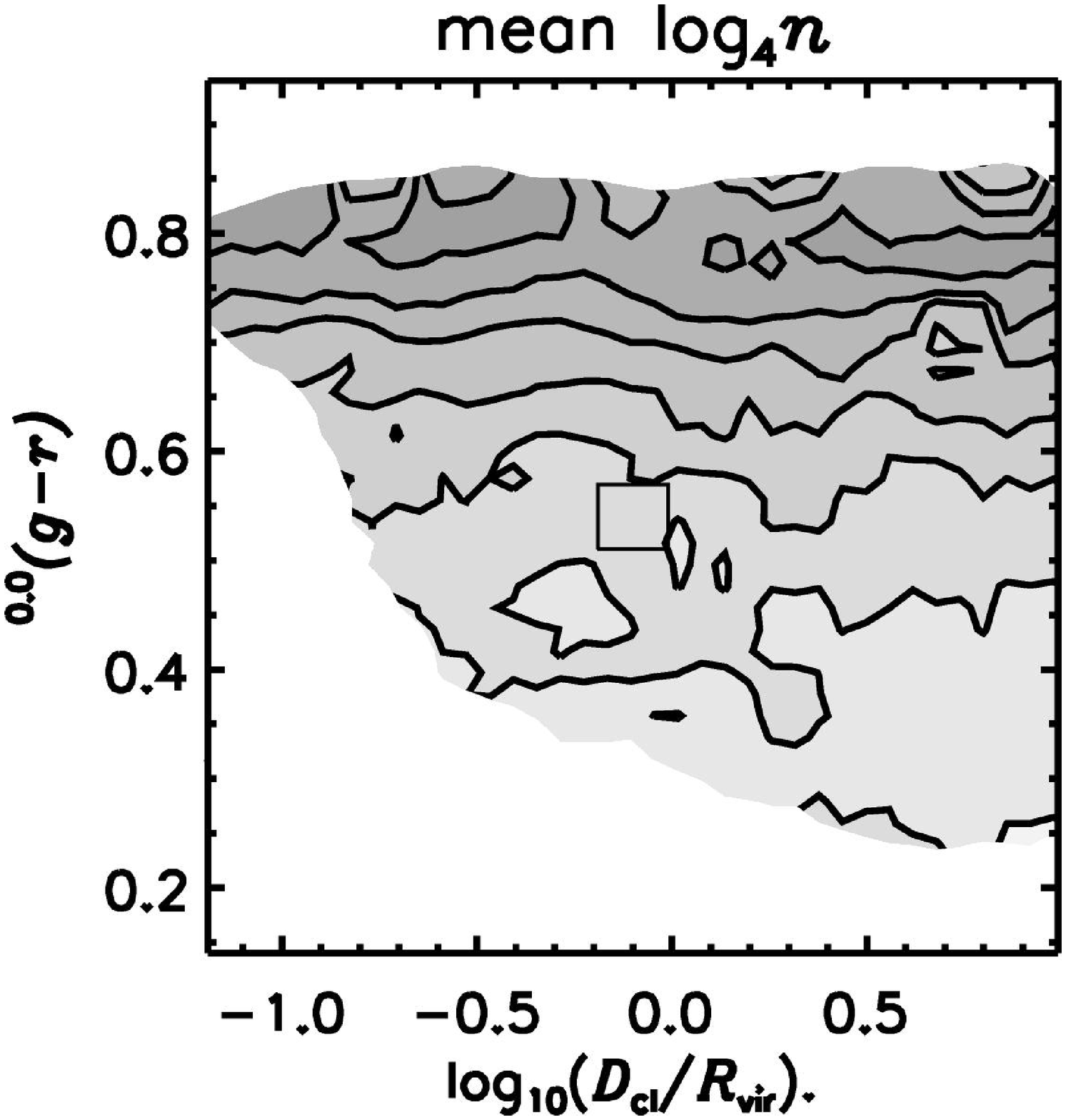}}~~~~%
\resizebox{2in}{!}{\includegraphics{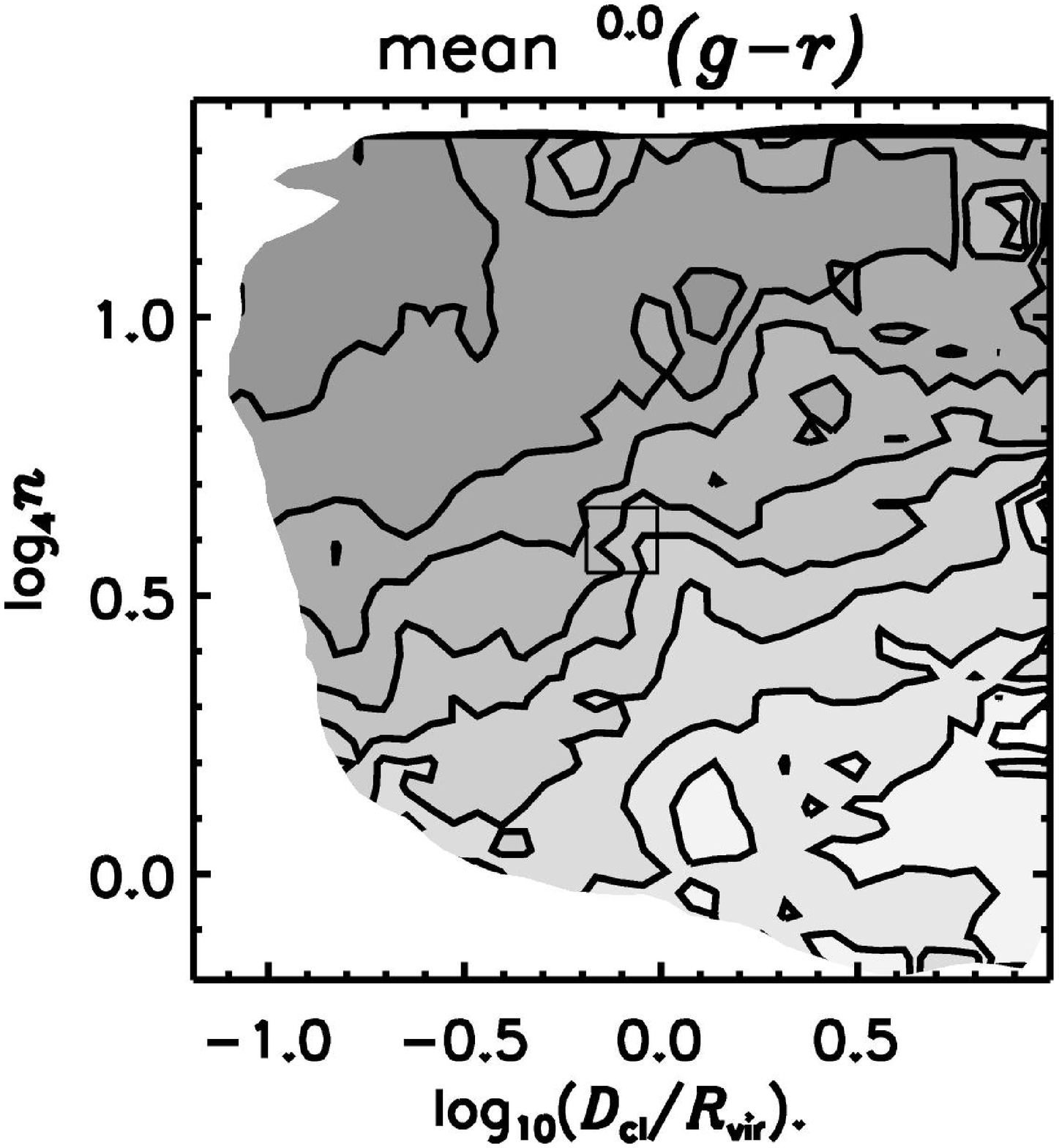}}~~~~%
\resizebox{2in}{!}{\includegraphics{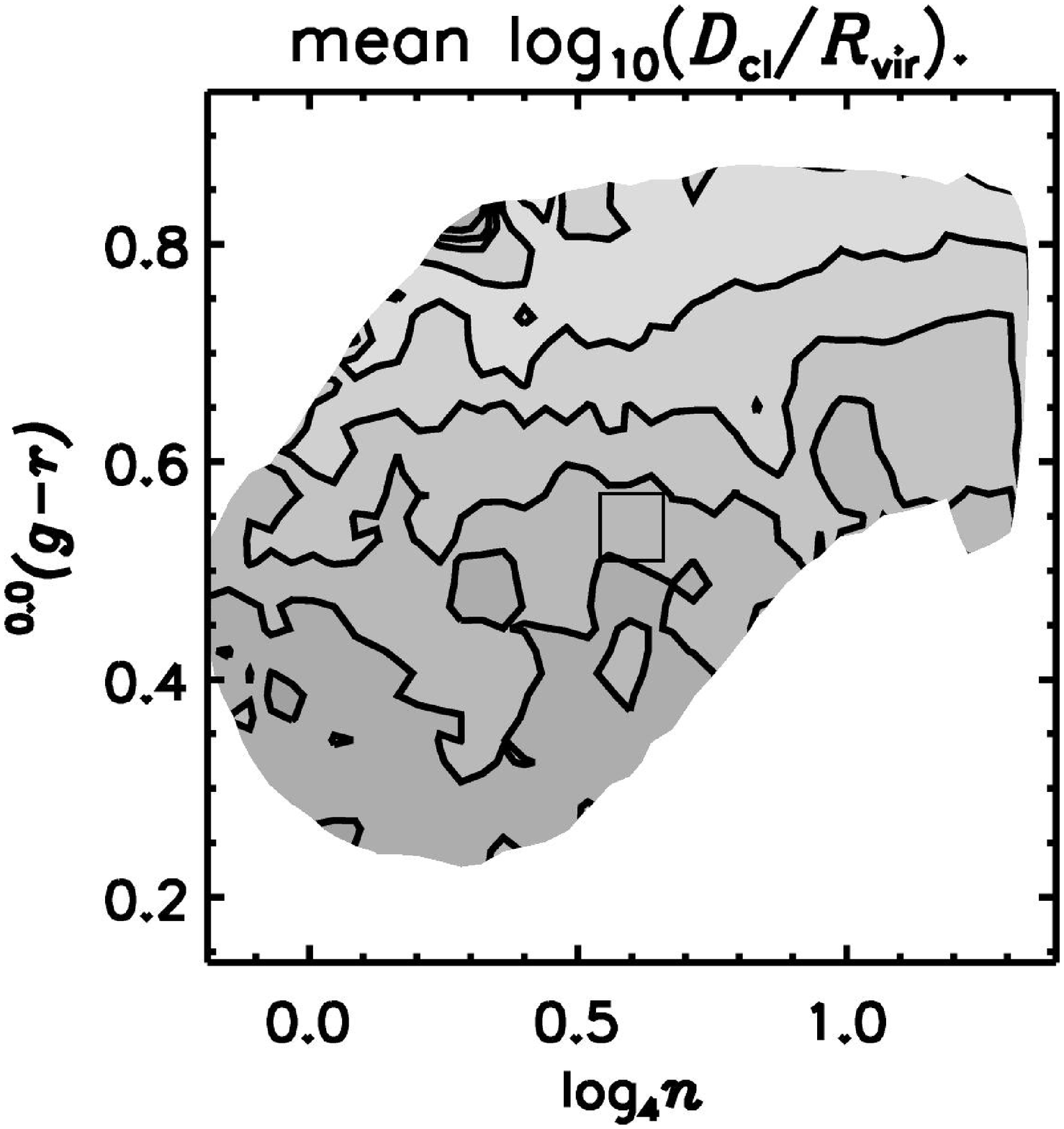}}
\end{center}
\caption{The left panel shows contours of mean radial concentration
$n$ as a function of $^{0.0}[g-r]$ color and clustocentric distance
\Dcl.  The mean is computed in a sliding box shown in the center of
the plot.  The center panel shows contours of mean color as a function
of concentration and clustocentric distance.  The right panel shows
contours of mean clustocentric distance as a function of color and
concentration.  Note that the contours in the left and right panels
are more horizontal than the contours in the center panel.
\label{fig:meanplots}}
\end{figure*}

Figure~\ref{fig:meanplots} displays the asymmetry shown in
Figure~\ref{fig:n_dont_matter} in a different way.  The left panel
shows that out of color and clustocentric distance, concentration
depends only on color since the mean concentration contours are
perpendicular to the color axis.  The center panel shows that color
depends on both concentration and clustocentric distance.  The right
panel shows that clustocentric distance depends more on color than
concentration, since the contours are nearly perpendicular to the
color axis.

\begin{figure*}
\begin{center}
\resizebox{!}{5.5in}{\includegraphics{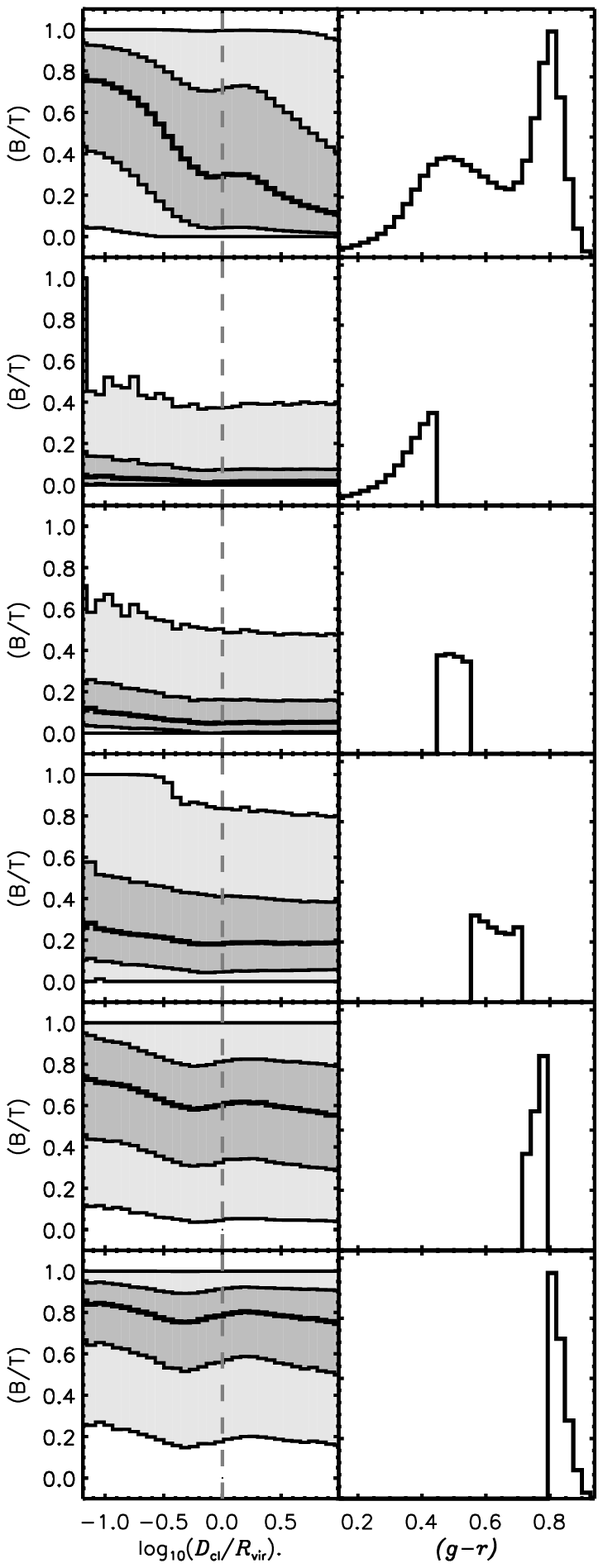}}%
\resizebox{!}{5.5in}{\includegraphics{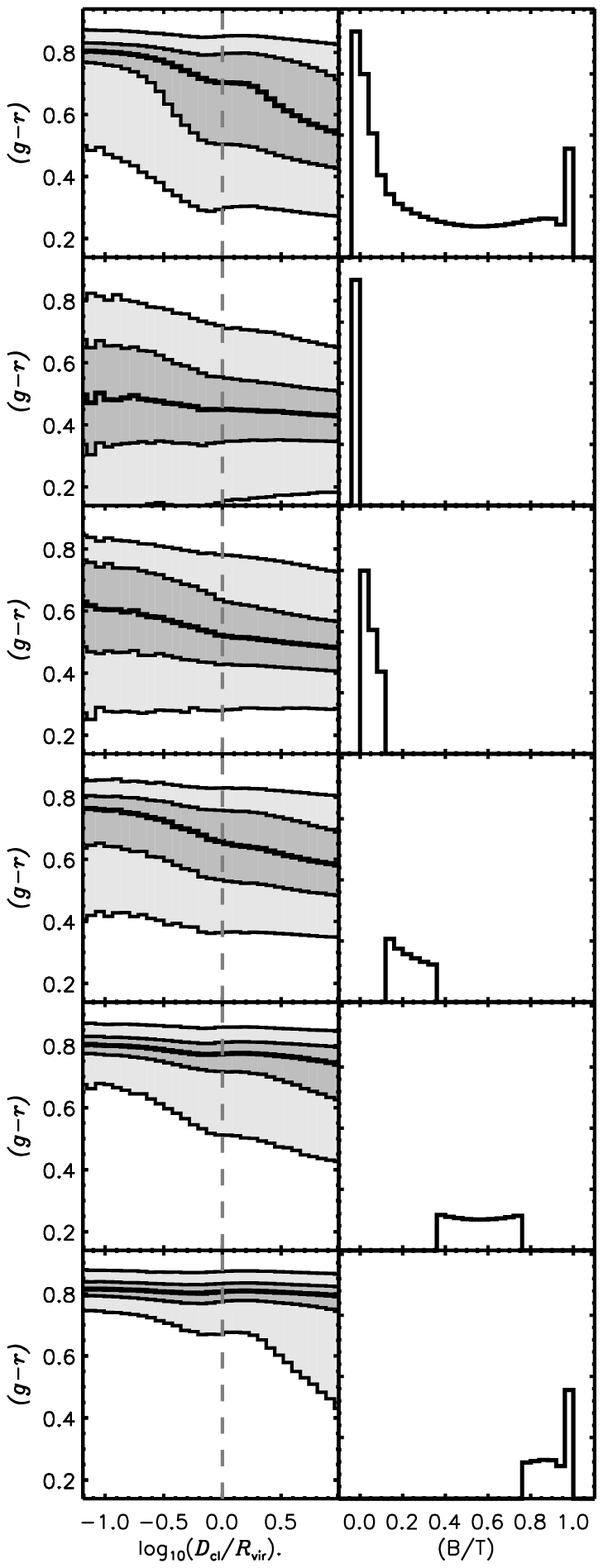}}
\end{center}
\caption{The same $2\times6$~grids shown in
Figure~\ref{fig:n_dont_matter} but showing the Millennium simulation
redshift zero outputs.  Instead of concentration $n$ we show the ratio $B/T$
of the bulge luminosity to total luminosity as a morphology indicator.  Notice
the asymmetry seen in Figure~\ref{fig:n_dont_matter} is not seen here.
\label{fig:n_dont_matter_mil}}
\end{figure*}

Figure~\ref{fig:n_dont_matter_mil} is analogous to
Figure~\ref{fig:n_dont_matter} but showing the redshift zero outputs
from the Millennium simulation.  Instead of concentration $n$ we show
the ratio of bulge luminosity to the total luminosity $B/T$.  Notice
that the asymmetry seen in the observed data is not reproduced here.
The asymetry is also absent when we replace color with star-formation
rate.

\begin{figure*}
\begin{center}
\resizebox{!}{5.5in}{\includegraphics{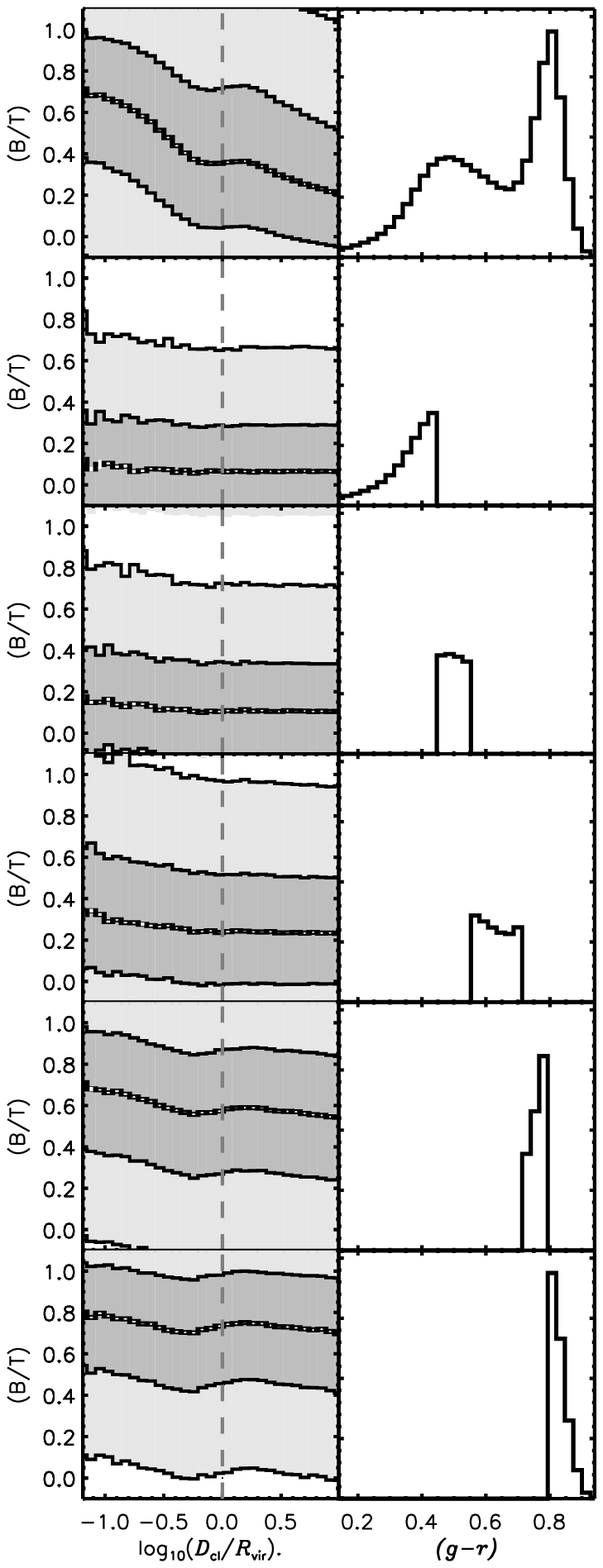}}%
\resizebox{!}{5.5in}{\includegraphics{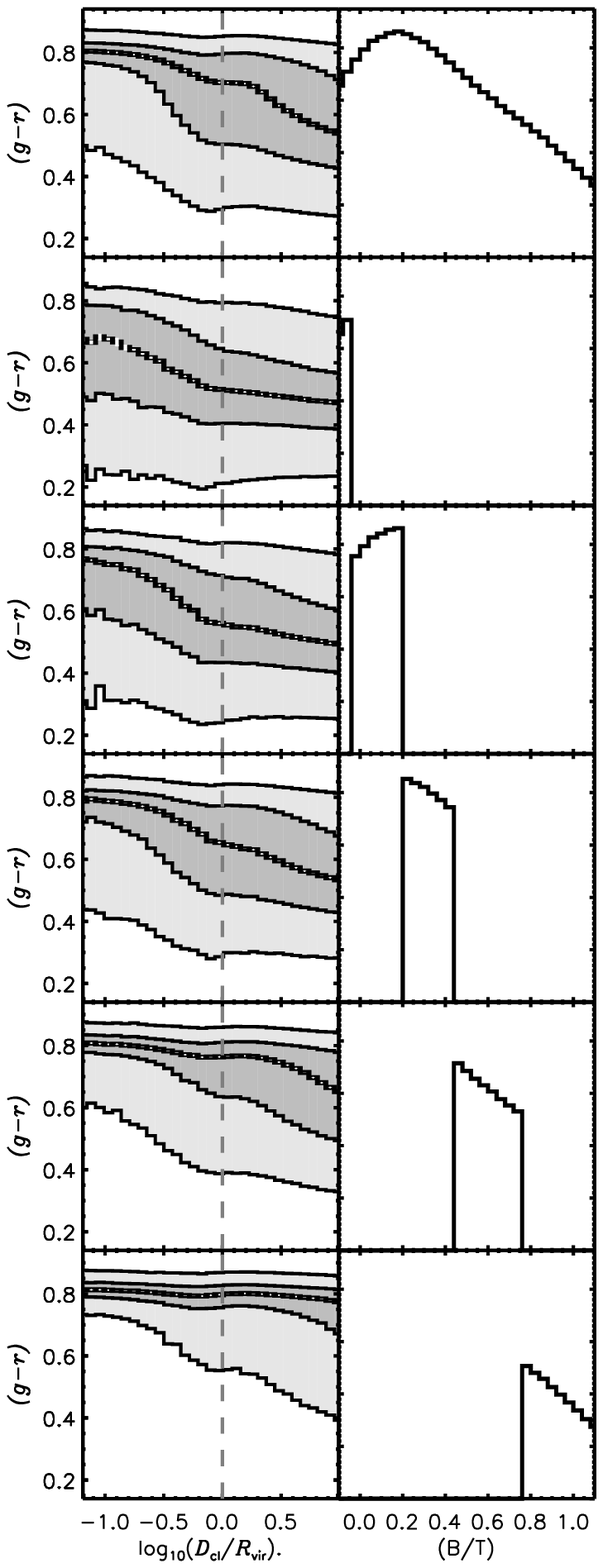}}
\end{center}
\caption{The same as Figure~\ref{fig:n_dont_matter_mil} but with
thirty percent scatter (see text) added to the bulge-to-total values.
Notice that the behavior of the trends do not significantly change
when adding these large errors to $B/T$.
\label{fig:btsig_n_dont_matter_mil}}
\end{figure*}

Figure~\ref{fig:btsig_n_dont_matter_mil} is the same as
Figure~\ref{fig:n_dont_matter_mil} but with 30 percent random errors
added to the $B/T$ values.  The ``random errors'' were produced by
randomly adding a Gaussian distribution with a dispursion of
$\sigma=0.3$ to the Millennium $B/T$ values. Notice that the behavior
of the trends do not significantly change when adding these large
errors to $B/T$.  This figure suggests that the asymmetry observed in
the SDSS data is not due to the fact that we measure star-formation
histories with much higher precision than we measure morphologies.

Another interesting result from Figures~1-3 is that the dependence of
star-formation-history and morphological parameters on clustocentric
distance is mostly limited to within the cluster virial radius.  The
top-left panels of these figures show that the trends observed at
distances of less than the virial radius (denoted by the vertical
dashed lines) nearly disappear at larger radii, where the relations
between star-formation/morphological parameters and clustocentric
distance are mostly flat.  This result is not reproduced by the
Millenium semi-analytic model.  The top-left panels of
Figure~\ref{fig:n_dont_matter_mil} show that, though the relations
between model galaxy parameters and clustocentric radius flatten out
near the virial radius, the strong trends resume at larger radii.

\section{Summary and Discussion}

Using a complete sample of 52,485 galaxies in the redshift range of
$0.015<z<0.068$ we examine how star-formation history
parameters---$^{0.0}[g-r]$ color and \Halpha\ EW---and morphology
parameters---radial concentration $n$ and surface brightness
$\mu_{^{0.0}i}$---relate to environment.  We use the transverse
projected distance to the nearest cluster---the clustocentric distance
\Dcl---as our environment parameter.  Similar to \cite{blanton05b}, we
find that our morphology tracers (concentration and surface
brightness) appear to be related to environment (\Dcl\, in this case)
only indirectly through their relationships with the star-formation
history tracers.  This suggests that the well known
morphology--environment relation is a result of the
star-formation-history--environment relation.  We note that, although
our simple morphology indicators do not show direct dependence on
clustocentric distance, this result does not disagree with that the
original morphology--density studies \citep{hubble36a, oemler74a,
dressler80a, postman84a} because these studies did not attempt to
separate morphological and star-formation dependences.  Our results
also do not disagree with previous work \citep{vogt04a} which suggests
that gas stripping plays a significant role in the morphological
transformation and rapid truncation of star formation by showing
asymmetries in HI and \Halpha\ flux on the leading edge of infalling
spiral galaxies.  There is no disagreement because our very blunt
morphology indicators are insensitive to those asymmetries, and the
asymmetries are only visible in a small fraction of all cluster
galaxies.

Where our results overlap those of previous investigators
\citep{christlein05a, blanton05b}, we mostly find good agreement.  In
particular, we find that galaxy properties depend on clustocentric
distance in much the same way as they do on other environmental
indicators \citep{blanton05b}.  Our result improves on this previous
one because the clustocentric distance is measured at much higher
signal-to-noise and probes environments on scales much smaller than
local overdensity measurements.  Our results improve on the findings
of \cite{christlein05a} by reproducing their result with far higher
signal to noise and by also by examining the inverse of their study.
They found that star formation rate is related to environment
independently of morphology.  We find that both color and star
formation rate behave in this way as well as that morphology doesn't
have a dependence on environment independent of star-formation
history.

It is worth noting that this asymmetry may, in principle, be
attributed to the \emph{measurements} that were made instead of the
properties of galaxies.  This is possible because our
star-formation-history properties are measured with far higher
precision than our morphology properties.  However, we do not believe
this is true for two main reasons; \emph{(1)} when we add large
asymmetric noise to the Millennium bulge-to-total luminosities we
still find no clear asymmetry and \emph{(2)} the fact that we observe
this asymmetry regardless of which star-formation history and
morphology tracers we use, strongly suggests that it is a description
of galaxy properties themselves.  We stress that our main result, that
there is no direct morphology--environment relation, holds for
morphology defined by concentration and surface brightness.  It
remains untested whether other characterizations of morphology, such
as spiral structure or asymmetry in the shape of the light
distribution, also show this trait.

Our results also show that, where there are trends between galaxy
properties and clustocentric distance, these trends are primarily
restricted to within the cluster virial radii.  This suggests that
environmental processes that affect galaxy properties are fairly local
(i.e., within the dark matter halo) and do not ``act at a distance''.
These results are not in agreement with those of \cite{gomez03a} and
\cite{lewis02a} who found that there are trends out to $\sim 3$ times
the virial radius.  The discrepancy can be attributed to our different
ways of estimating cluster virial radii.  These authors use the
\cite{girardi98a} approximation relating virial radius to velocity
dispersion, whereas we use virial radii based on the \cite{berlind06a}
mass estimates, which are derived from cluster abundances.  We have
checked the \cite{girardi98a} approximation on our clusters and
find that it yields virial radii that are $\sim 2$ times smaller than
the ones we use, which explains much of the discrepancy.  We trust our
mass estimates because they are based on abundances, which have been
shown to be unbiased for our clusters \cite{berlind06a}.  Masses
based on the \cite{girardi98a} approximation yield an incorrect
cluster mass function.

The asymmetry that we see between star-formation and morphological
parameters has interesting implications for the role that environment
plays in shaping galaxy properties.  Our results suggest that
environmental processes can affect galaxy star-formation histories
without simultaneously affecting morphology, but in cases where
an environmental mechanism does affect galaxy morphology, it must
also affect star-formation history.   Another possible explanation
is that environment affects both star-formation history and morphology,
but the changes in star-formation history occur on a much more rapid
timescale than the corresponding changes to morphology.

In this short paper, we have investigated how galaxy star-formation
histories and morphologies depend on environment.  We find the
following asymmetry: At fixed color, there is no dependence of
morphology on environment, while at fixed morphology, there is a
residual dependence of color on environment.  This result suggests
that the morphology--environment relation is a result of the
combination of the star-formation-history--environment and the
morphology--star-formation-history relations.  Moreover, we performed
a similar analysis on a semi-analytic model of galaxy formation and
find that this asymmetry is not reproduced.  Our results have
important implications for the role that environment plays in shaping
galaxy star-formation histories and morphology, and they place strong
constraints on models of galaxy formation.

\acknowledgements We thank Alison Coil, Alister Graham, Erin Sheldon,
and Beth Willman for useful ideas, conversations, and comments on the
manuscript.  The Millennium Run simulation used in this paper was
carried out by the Virgo Supercomputing Consortium at the Computing
Centre of the Max-Planck Society in Garching. The semi-analytic galaxy
catalogue is publicly available at
http://www.mpa-garching.mpg.de/galform/agnpaper.  ADQ and DWH thank
Hans-Walter Rix and the Max-Planck-Institut f\"ur Astronomie for
hospitality.  This research made use of the NASA Astrophysics Data
System.  ADQ, DWH, and MRB are partially supported by NASA (grant
NAG5-11669) and NSF (grant AST-0428465).

Funding for the creation and distribution of the SDSS Archive has been
provided by the Alfred P. Sloan Foundation, the Participating
Institutions, the National Aeronautics and Space Administration, the
National Science Foundation, the U.S. Department of Energy, the
Japanese Monbukagakusho, and the Max Planck Society. The SDSS Web site
is http://www.sdss.org/.

The SDSS is managed by the Astrophysical Research Consortium for the
Participating Institutions. The Participating Institutions are The
University of Chicago, Fermilab, the Institute for Advanced Study, the
Japan Participation Group, The Johns Hopkins University, Los Alamos
National Laboratory, the Max-Planck-Institute for Astronomy, the
Max-Planck-Institute for Astrophysics, New Mexico State University,
University of Pittsburgh, Princeton University, the United States
Naval Observatory, and the University of Washington.

\clearpage

\clearpage

\clearpage

\clearpage

\clearpage

\clearpage

\end{document}